\documentclass[amsmath,10pt,twocolumn,superscriptaddress,footinbib,pre,pre]{revtex4-2}
\usepackage{amsmath,amsfonts,bm,dsfont}
\usepackage{graphicx}
\usepackage{subfigure}
\usepackage{float}
\usepackage{chemformula}
\usepackage[normalem]{ulem}
\usepackage{physics}
\usepackage{uri}

\usepackage{color}
\usepackage[colorlinks=true,allcolors=blue]{hyperref}
\usepackage{tikz}
\usepackage{upgreek}
\usepackage{amssymb}

\DeclareUnicodeCharacter{0301}{\'{e}}

\newcommand{\bx}{{\bf x}}

\begin{document}
\title{From high-dimensional committors to reactive insights}
\author{Nils E.~Strand}
\altaffiliation[Now at ]{James Franck Institute, University of Chicago, Chicago, IL 60637, USA}
\author{Schuyler B.~Nicholson}
\author{Hadrien Vroylandt}
\altaffiliation[Now at ]{CERMICS, {\'E}cole des Ponts ParisTech.}
\author{Todd R.~Gingrich}
\email{todd.gingrich@northwestern.edu.}
\affiliation{Department of Chemistry, Northwestern University, 2145 Sheridan Road, Evanston, Illinois 60208, USA}

\begin{abstract}
  Transition path theory (TPT) offers a powerful formalism for extracting the rate and mechanism of rare dynamical transitions between metastable states.
  Most applications of TPT either focus on systems with modestly sized state spaces or use collective variables to try to tame the curse of dimensionality.
  Increasingly, expressive function approximators like neural networks and tensor networks have shown promise in computing the central object of TPT, the committor function, even in very high dimensional systems.
  That progress prompts our consideration of how one could use such a high dimensional function to extract mechanistic insight.
  Here, we present and illustrate a straightforward but powerful way to track how individual dynamical coordinates evolve during a reactive event.
  The strategy, which involves marginalizing the reactive ensemble, naturally captures the evolution of the dynamical coordinate's distribution, not just its mean reactive behavior.

\end{abstract}
\maketitle
\section{Introduction}

Biophysical systems frequently involve dynamics that is both high-dimensional and stochastic~\cite{Svoboda1993, McAdams1997, Zhuang2000, Eaton2000, Golding2005, allen2005sampling, Eldar2010}.
When those dynamical processes relax into an equilibrium, it is possible to study the stable states in terms of thermodynamics without reference to the dynamics.
Many biophysical processes, however, operate away from equilibrium, and in that regime, it is especially crucial to understand the kinetics.
Studying Markovian models is a well-developed route to analyzing that high-dimensional stochastic kinetics, a route common to the chemical master equation~\cite{gillespie1992rigorous}, Langevin dynamics~\cite{vankampen1992}, and Markov State Models (MSMs)~\cite{chodera2014markov,husic2018markov}.
It is often the case that those Markovian models exhibit slow-timescale transitions between different regions of a configuration space, transitions one would associate with barrier crossing in an equilibrium setting.
Owing to the chemical physics history, we generically refer to switches from one set of metastable configurations \(\mathcal A\) into another \(\mathcal{B}\) as a reaction or reactive event.
Examples of such reactions are transitions between metastable states in gene regulatory networks and chemical reaction networks~\cite{thomas2014phenotypic,graciun2006understanding,ozbudak2004multistability}.
Because the configuration space can become astronomically vast, one often seeks a coarse-grained description of the kinetics: what are the long-lived metastable regions of configuration space, what are the timescales for reactions, and what is the mechanism of the reaction?
That mechanism is particularly desirable since it is easier to design ways to modify the reaction rate if one knows \emph{how} the reaction typically proceeds.

The most straightforward approach to learning the mechanism involves generating and watching ensembles of representative reactive trajectories~\cite{gillespie2013perspective,yang2006efficient} to form impressions of how those representative trajectories progress from \(\mathcal A\) to \(\mathcal B\).
Due to a separation of timescales between the typical residence time in metastable states and the transition time~\cite{hanggi1990reaction}, it can be impractical to directly simulate and watch the large number of required trajectories.
Enhanced sampling methods like transition path sampling (TPS)~\cite{dellago1998transition,bolhuis2002throwing,crooks2001efficient,gingrich2015preserving} and forward flux sampling (FFS)~\cite{allen2005sampling,allen2006simulating,allen2009forward} can offer more efficient ways to generate the ensemble of reactive trajectories, but even when the ensemble can be sampled, the results are still high-dimensional making it non-trivial to interpret. What is needed is a low-dimensional representation of the resulting mechanism from the high-dimensional reactive trajectories.

To circumvent these challenges, one can parameterize progress not by time but by a one-dimensional reaction coordinate.
The reaction coordinate can be thought of as a many-to-one mapping from microstate \({\bf x}\) onto a single variable measuring progress along the reaction, and the best choice for such a progress coordinate is known to be the so-called committor function, \(q({\bf x})\)~\cite{weinan2010transition, lu2014exact}.
Transition path theory (TPT)~\cite{weinan2006towards,metzner2006illustration,metzner2009transition,weinan2010transition} provides explicit expressions to compute \(q({\bf x})\) in terms of a generator of the dynamics, but the cost of directly performing such computations rapidly increases with the number of microstates.
Due to the curse of dimensionality, it is common for problems of interest to have astronomically many microstates.
In those cases, the most common way committor functions have been used for complex systems is to avoid \(q({\bf x})\) and instead compute a committor function defined over low-dimensional (often one-dimensional) collective variable \({\bf y}\), which is a function of \({\bf x}\).
This committor \(q({\bf y})\) is practically computed by sampling.
For example, many trajectories can be initialized with a particular value of \({\bf y}\) then propagated until they reach either \(\mathcal{A}\) or \(\mathcal{B}\), with \(q({\bf y})\) being the probability of first reaching \(\mathcal{B}\).
Approaches built around committors of one or more collective variables have been productive~\cite{du1998transition,bolhuis2000reaction,rhee2005one,ma2005,lechner2010nonlinear, rohrdanz2013discovering, neupane2016protein, elber2017calculating,he2022committor,roux2022transition,jung2023machine,evans2023computing}, but the approaches typically require choosing good collective variables up front.
A significant body of research has developed strategies for identifying and optimizing those ``good'' collective variables, ideally finding a \({\bf y}\) that resembles the committor itself~\cite{geissler1999kinetic,bolhuis2002throwing,hummer2004transition,best2005reaction,banushkina2016optimal,peters2016reaction,zhang2016effective,krivov2018protein,wu2022rigorous,mouaffac2023optimal,chen2023discovering}.
More recently, basis expansions~\cite{thiede2019galerkin,strahan2021long,aristoff2024fast,guo2024dynamics}, neural networks~\cite{rotskoff2022active,strahan2023predicting,ma2005,khoo2018solving,hasyim2022supervised,li2019computing,jacques2023data,lin2024deep} and tensor networks~\cite{chen2023committor} have been used to estimate \(q({\bf y})\) from sampled trajectories even when \({\bf y}\) is quite high dimensional.
  Those advances pair nicely with strategies to extract the mechanism of the reactive events in the collective variable space from \(q({\bf y})\)~\cite{noe2009constructing,lane2011markov,finkel2023data}.
  In particular, Ref.~\cite{guo2024dynamics} has used that \(q({\bf y})\) to inspect how the steady-state distribution of certain collective variables varies as a function of reaction progress.

Suppose, by contrast, one could discard the collective variables altogether and it were practical to solve for the full-dimensional committor \(q({\bf x})\).
Here, we introduce and illustrate that for discrete systems like well-mixed chemical reaction networks, it is indeed numerically practical to compute \(q({\bf x})\) and extract mechanistic insight.
For those problems, the strategy does not even require trajectory sampling.
The key idea of this paper is that access to the full-dimensional committor \(q({\bf x})\) allows one to inspect how each dynamical coordinate \(x_i \in {\bf x}\) evolves as a function of reaction progress.
Crucially, this approach retains a distribution over \(x_i\), not just the mean behavior, allowing the approach to directly reveal the presence of multiple reactive pathways.

The methodology is built upon TPT's reactive ensemble, which gives the density \(\rho^{\mathcal{AB}}({\bf x})\) of occupying a microstate \({\bf x}\) given that that the system is in the midst of transitioning from \(\mathcal{A}\) to \(\mathcal{B}\).
For each degree of freedom \(x_i\), \(1\leq i\leq D\), we compute a two-dimensional distribution formed from the reactive ensemble by marginalizing over all other degrees of freedom:
 \begin{align}
   \nonumber \rho^{\mathcal{AB}}&(x_i, q)\\
   &= \int dx_1 \cdots dx_{i-1} dx_{i+1}\cdots dx_D \rho^{\mathcal{AB}}({\bf x}) \delta(q({\bf x}) - q),
   \label{eq:main}
 \end{align}
 with the \(\delta\) function serving to pick out how far the reaction had progressed toward \(\mathcal{B}\).
 The marginal \(\rho^{\mathcal{AB}}(x_i, q)\) highlights the single coordinate \(x_i\), but it retains the influence of the other coordinates only in so far as they impact the progress coordinate \(q\).
 In this way, one can view how the distribution for each \(x_i\) evolves during the reaction process, parameterized by \(q\).
 The approach has the flavor of the so-called violin plots of Ref.~\cite{guo2024dynamics}, but it computes the \(q\) dependence of the reactive, rather than steady-state, ensemble.
 The reactive ensemble might be a preferred choice if the steady-state has an overwhelming amount of density within the regions \(\mathcal A\) and \(\mathcal B\), causing probabilities separating the two metastable basins to be difficult to visualize.
 
We illustrate the idea with two example problems, both of which admit a direct computation of \(q\) over a discrete state space.
First, we demonstrate the approach for a discretized two-dimensional (2D) diffusion problem where the explicit calculation of the committor has been previously studied~\cite{metzner2009transition, cameron2014flows,louwerse2022information}.
Though only two dimensional, this problem illustrates the approach and emphasizes that it can naturally highlight when multiple distinct pathways meaningfully contribute to reactive events.
Second, we move to a situation with too many degrees of freedom to straightforwardly plot \(q({\bf x})\), a gene toggle switch (GTS) model~\cite{allen2005sampling} with two metastable states emerging from stochastic chemical kinetics of seven chemical species.
Calculating the committor for the GTS model is more complicated than most literature toy problems since we consider a GTS model with several million microstates, many more than coarse-grained models typically used for transition path analysis~\cite{noe2009constructing,voelz2009molecular,shukla2014activation,guamera2016optimized,liu2019markov,banerjee2014transition,vani2022computing}.
Using sparse linear algebra methods, we compute \(q\) and show how it can be used to extract the reaction mechanism one species at a time.

\section{Transition path theory}
\subsection{Standard formulation}

Our work builds upon TPT, so we start by reviewing its main results for continuous-time, discrete-state Markov dynamics~\cite{weinan2010transition}.
One can choose a canonical ordering of microstates so a many-body microstate \(\bx\) is labeled by the single index \(i\).
Let \(W_{ij}\) denote the rate, or probability per unit time, of transitions from the \(j^{\rm th}\) into the \(i^{\rm th}\) microstate.
Conservation of probability is enforced because the diagonal elements \(W_{ii}\) are chosen such that \(\sum_i W_{ij} = 0\).
Without loss of generality, we assume it is possible to reach any microstate from any other microstate in a finite number of transitions, that is to say \(W\) is irreducible.
In the long time limit, microstate \(i\) is visited with steady-state probability \(\pi_i\).
That distribution follows simply from the matrix \(W\) as the solution to
\begin{equation}
W{\boldsymbol\pi}=0,
\label{eq:ss}
\end{equation}
where \({\boldsymbol \pi}\) is the vector of steady state probabilities for each microstate.

  TPT partitions the space of microstates \(\mathcal{S}\) into three regions: \(\mathcal{A}, \mathcal{B},\) and \((\mathcal{A} \cup \mathcal{B})^{\rm c}\), where the superscript c is the complement of the set.
The aim is to describe properties of the Markov dynamics within \((\mathcal{A} \cup \mathcal{B})^{\rm c}\) conditioned upon starting in \(\mathcal{A}\) and ending in \(\mathcal{B}\), without having first returned to \(\mathcal{A}\).
This conditioned process is of special physical interest when \(\mathcal{A}\) and \(\mathcal{B}\) are metastable states and trajectories pass through \((\mathcal{A} \cup \mathcal{B})^{\rm c}\) rarely.
Then the rare transitions are viewed as reactions.
A motivating goal for TPT was to compute the reaction rate \(k_{\mathcal{A} \mathcal{B}}\) from the Markov rate operator \(W\).
It has been shown that this reaction rate is expressed compactly in terms of the committor function, specifically the forward committor function, which we distinguish with a superscript +.
For the discrete state space, we define the vector \({\bf q^+}\) whose element \(q^+_i\) is the probability a trajectory initiated in state \(i\) will reach \(\mathcal{B}\) before \(\mathcal{A}\).
The forward committor solves the Dirichlet boundary value problem
\begin{align}
    \begin{cases}\sum_{j \in \mathcal{S}}q^+_j W_{ji}=0,\quad&\forall\, i \in(\mathcal A\cup\mathcal B)^{\rm c}\\
    q^+_i=0,\quad&\forall\, i \in\mathcal A\\
    q^+_i=1,\quad&\forall\, i \in\mathcal B\end{cases}.
\end{align}
Practically, it is convenient to cast that problem as the linear equation
\begin{equation}
    U{\bf q^+}={\bf v},
    \label{eq:forward}
\end{equation}
where \(U\) is a square matrix with elements \(U_{ji}=W_{ij}\), for all \(i,j \notin\mathcal A\cup\mathcal B\), and \({\bf v}\) is a vector with elements \(v_j = - \sum_{i \in\mathcal B} W_{ij}\), for all \(j\in(\mathcal A\cup\mathcal B)^{\rm c}\), and zero otherwise.
In words, \(U\) is the transpose of the submatrix of \(W\) corresponding to the reactive region, that is, the sites not in \(\mathcal A\) or \(\mathcal B\).
Multiplying the forward committor vector \(\bf q^+\) from the left with \(U\) results in \(\bf v\), a vector whose element \(i\notin\mathcal A\cup\mathcal B\) is minus the sum of the rates leaving \(i\) and entering \(\mathcal B\).

TPT defines the backward committor \({\bf q^-}\) in a manner analogous to \({\bf q^+}\); \(q_i^-\) is the probability of being at \(i\) given that the system last occupied \(\mathcal{A}\) before \(\mathcal{B}\).
The backward committor relies on the time-reversed process, characterized by a rate matrix \(\tilde W\).
The off-diagonal elements of this matrix are given by \(\tilde W_{ij} = W_{ji}\pi_i/\pi_j\), and diagonal elements, \(\tilde W_{ii}=-\sum_{j\neq i}\tilde W_{ji}\)~\cite{metzner2009transition}.
The boundary value problem for the backward committor,
\begin{align}
    \begin{cases}\sum_{j \in\mathcal S}q_j^-\tilde W_{ji}=0,\quad&\forall\, i\in (\mathcal A\cup\mathcal B)^{\rm c}\\
    q_i^-=1,\quad&\forall\, \bx \in\mathcal A\\
    q_i^-=0,\quad&\forall\, \bx \in\mathcal B\end{cases},
\end{align}
leads to the linear equation
\begin{equation}
    \tilde U{\bf q^-}=\tilde {\bf v},
    \label{eq:backward}
\end{equation}
where \(\tilde U_{ij}=\tilde W_{ji}\), for all \(i,j\notin\mathcal A\cup\mathcal B\) and \(\tilde v_i=-\sum_{k\in\mathcal A}\tilde W_{ki}\), for all \(i\in (\mathcal A\cup\mathcal B)^{\rm c}\).
For reversible systems, \(W=\tilde W\), and the forward and backward committors are trivially related as \(q_i^-=1-q_i^+\).
Calculations in this paper involve Markov dynamics with a \(W\) that breaks detailed balance, requiring Eqs.~\eqref{eq:forward} and \eqref{eq:backward} to be independently solved.
By additionally solving for the steady-state distribution \({\boldsymbol \pi}\), one can then construct the reactive probability
\begin{equation}
    P_i^{\mathcal{AB}}=\pi_iq_i^+q_i^-.
    \label{eq:mAB}
\end{equation}
This \(P_i^{\mathcal{AB}}\) is the probability that a reactive trajectory occupies discrete microstate \(i\)~\cite{metzner2009transition}.
If those microstates come from a discretization of a continuous problem, then the reactive density \(\rho^{\mathcal{AB}}(\bx) = P_i^{\mathcal{AB}} / V\), where \(i\) is the index for microstate \(\bx\) and \(V\) is the volume element of each discretized cell.

\subsection{Re-expression for large-scale systems breaking detailed balance}
\label{sec:reexpression}

When detailed balance is not satisfied, as in our second example problem, \({\bf q^-}\) does not follow directly from \({\bf q^+}\).
In those cases, \({\bf q^-}\) could in principle be found by solving Eq.~\eqref{eq:backward}, but that approach is impractical for large systems.
Elements of the time-reversed rate matrix can suffer from numerical instability due to states with vanishingly small steady-state probabilities entering into the denominator of \(\tilde{W}_{ij} = W_{ji} \pi_i / \pi_j\).
Note, however, that we can directly solve for the vector \({\bf r}\) whose elements \(r_i = \pi_i q_i^-\) appear in the expression for the reactive density, Eq.~\eqref{eq:mAB}.
In Appendix~\ref{sec:committors}, we demonstrate that solving Eq.~\ref{eq:backward} is equivalent to solving the linear equation
\begin{equation}
    U^T {\bf r}={\bf s},
    \label{eq:backward2}
\end{equation}
where \(s_i=-\sum_{k\in\mathcal A}W_{ik}\pi_k\) for all \(i\in(\mathcal A\cup\mathcal B)^{\rm c}\).
Notice that unlike Eq.~\eqref{eq:backward}, Eq.~\eqref{eq:backward2} requires a transpose of the operator \(U\) used to solve the forward committor, not the numerically problematic time-reversal.
The transformation doesn't give a free lunch in that the vector \({\bf s}\) cannot be constructed as simply as \(\mathbf{v}\) and \(\mathbf{\tilde{v}}\) were.
Instead, \({\bf s}\) requires knowledge of the steady state \(\boldsymbol \pi\).
We therefore solve for \({\bf r}\) in two stages.
First, we find \(\boldsymbol \pi\) by applying Arnoldi iteration to the eigenvalue problem in Eq.~\eqref{eq:ss}.
With that \(\boldsymbol \pi\), we construct \({\bf s}\) and use general minimum residual (GMRES) iterations to solve Eqs.~\eqref{eq:forward} and~\eqref{eq:backward2} for \({\bf q^+}\) and \({\bf r}\).
Combining the two, we obtain the reactive distribution as \(P^{\mathcal{AB}}_i = q^+_i r_i\).

\section{Results}
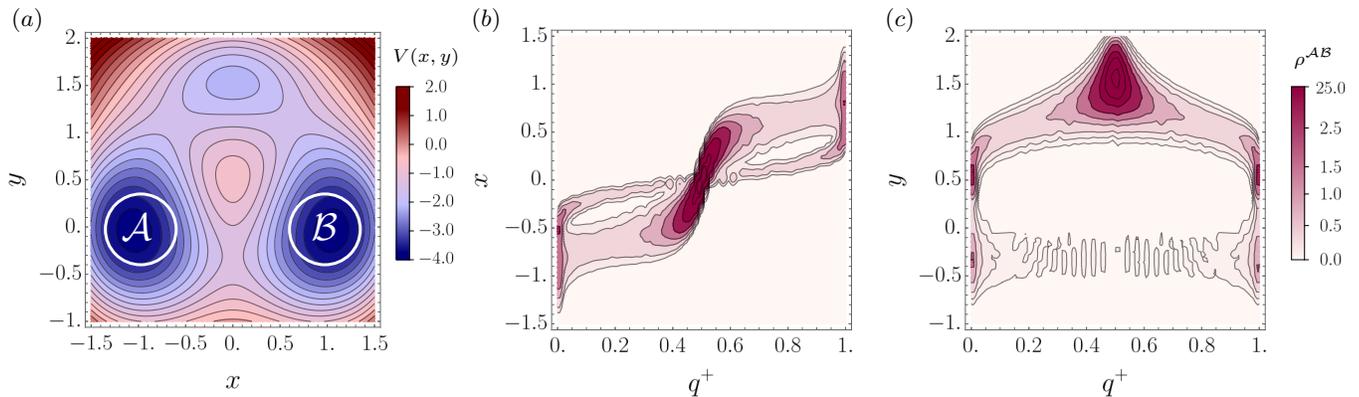
\begin{figure*}
    \centering
    \begin{tikzpicture}
        \node at (-2.5,0) {\includegraphics[width=0.29\textwidth]{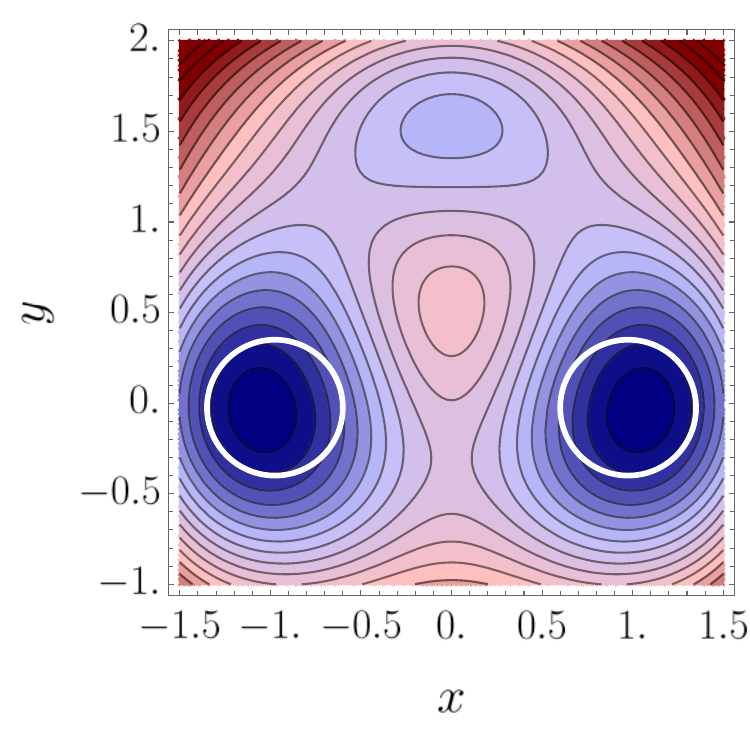}};
        \node at (3.7,0) {\includegraphics[width=0.29\textwidth]{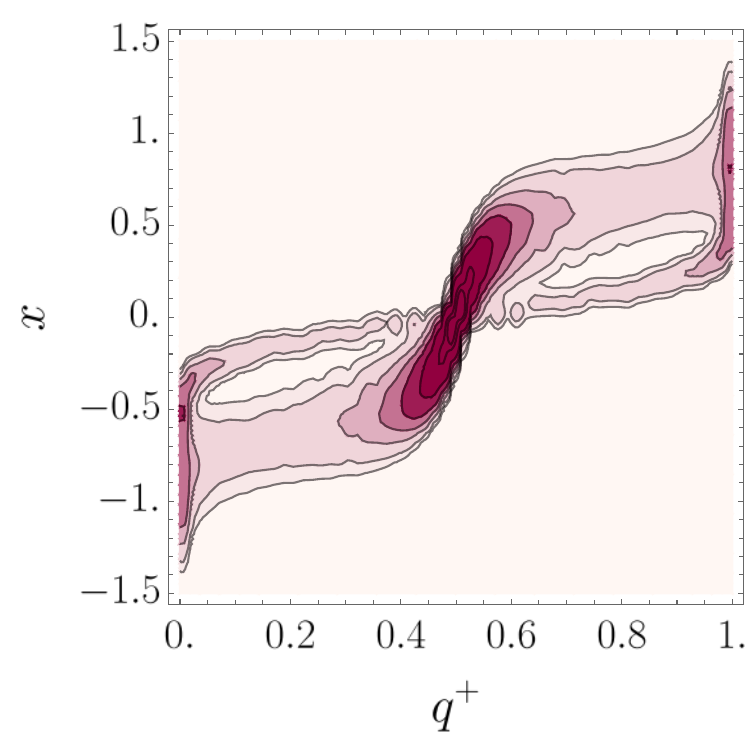}};
        \node at (9.2,0) {\includegraphics[width=0.29\textwidth]{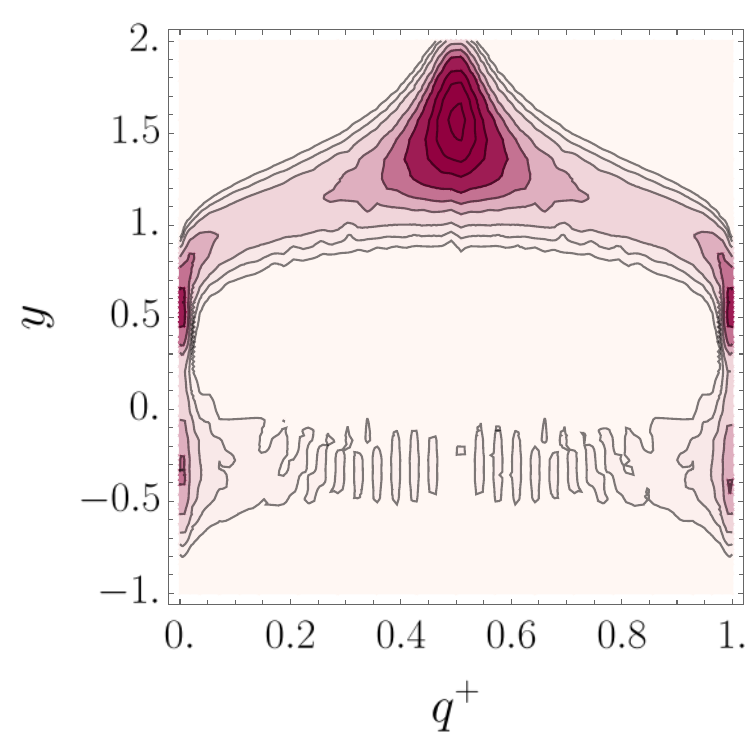}};
        \node at (0.8,0.5) {\includegraphics[width=0.08\textwidth]{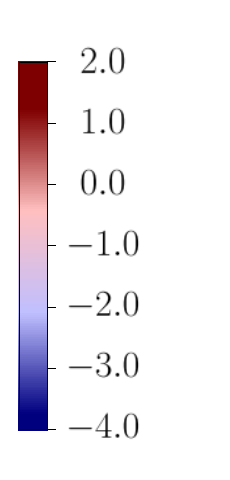}};
        \node at (12.7,0.5) {\includegraphics[width=0.08\textwidth]{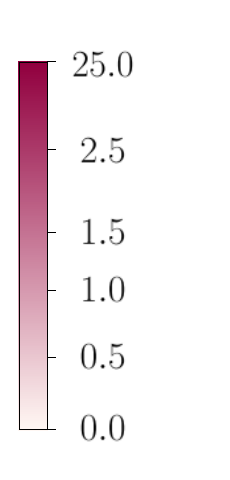}};
        \node at (0.6,2) {\scriptsize\(V(x,y)\)};
        \node at (12.4,2) {\scriptsize\(\rho^{\mathcal{AB}}\)};
        \node (a) at (-4.7,2.5) {\((a)\)};
        \node (a) at (1.4,2.5) {\((b)\)};
        \node (a) at (6.9,2.5) {\((c)\)};
        \node at (-3.25,-0.25) {\color{white}\Large\(\mathcal{A}\)};
        \node at (-0.75,-0.25) {\color{white}\Large\(\mathcal{B}\)};
    \end{tikzpicture}
    \caption{\textbf{Detail-balanced dynamics of a single thermal particles.}
      Following Ref.~\cite{metzner2009transition}, we study overdamped dynamics on an energy landscape with two deep wells and seek information about the mechanism passing from reactants \(\mathcal{A}\) to products \(\mathcal{B}\).
      Upon discretizing the continuous state space on a grid, sparse linear algebra methods give \(\boldsymbol \pi\), \({\bf q^+}\), and \({\bf q^-} = 1 - {\bf q^+}\) for that discretized problem.
      The reactive density \(\rho^{\mathcal{AB}}(\mathbf{x})\) can be marginalized as in Eq.~\eqref{eq:main} to reveal the distribution of each coordinate (\(x\) and \(y\)) as a function of reaction progress \(q^+\).
      Plots of \(\rho^{\mathcal{AB}}(x, q^+)\) and \(\rho^{\mathcal{AB}}(y, q^+)\) reflect the distribution of outcomes for coordinate \(x\) and \(y\), respectively, where the other coordinate's state is considered only to the extent that it impacts the reaction progress \(q^+\).
        Retaining information about the statistical ensemble naturally reveals the presence of multiple reactive pathways.
        For cleaner visualization, the reactive ensemble densities were smoothed with Gaussian kernels and bandwidths obeying Silverman's rule~\cite{silverman1986} and colored with a nonlinear hyperbolic tangent colorbar that enhances the resolution of the low-probability regions.
    }
    \label{fig:2ddiffusion}
\end{figure*}

\subsection{Two-dimensional diffusion on a metastable landscape}
\label{sec:2ddiffusion}

Before breaking detailed balance or considering high-dimensional systems, it is useful to discuss a simpler prototypical minimal example, that of two-dimensional diffusion on a metastable landscape~\cite{metzner2009transition}.
For this example, the microstates \(\mathbf{x}\) are defined by two coordinates, \(x\) and \(y\).
The system evolves on the three-well energy landscape
\begin{align}
\nonumber  V(x,y) &= 3e^{-x^2-\left(y-\frac{1}{3}\right)^2}-3e^{-x^2-\left(y-\frac{5}{3}\right)^2}\\
\nonumber  & \ \ - 5e^{-(x-1)^2-y^2}-5e^{-(x+1)^2-y^2}\\
  & \ \ + \frac{x^4}{5}+\frac{(y-\frac{1}{3})^4}{5},
  \label{eq:landscape}
\end{align}
according to an overdamped Langevin dynamics with a gradient force and a random force \(\boldsymbol{\xi}\) of thermal origin:
\begin{equation}
  \dot{\mathbf{x}}=-\nabla V(\mathbf{x}) + \boldsymbol{\xi}.
  \label{eq:langevin}
\end{equation}
With inverse temperature \(\beta\), the white noise satisfies \(\langle\xi_i(t)\xi_j(t')\rangle=2\beta^{-1}\delta_{ij}\delta(t-t')\).
The energy landscape was constructed to have metastable basins (see Fig.~\ref{fig:2ddiffusion}a), and the standard problem is to describe the rare dynamical path that causes the system to transition from one of those basins to the other.
This particular toy problem is a useful starting point because the state space can be discretized onto a 200 \(\times\) 200 grid such that the corresponding linear equation, Eq.~\eqref{eq:forward}, for the forward committor can be solved~\cite{metzner2009transition}.
With an only two-dimensional \(\mathbf{x}\), the solution \(q^+(\mathbf{x})\) can be plotted to offer a clear visual for how the reactions proceed from one basin to the other.
The landscape has two distinct pathways along which transitions can occur, and a plot of \(q^+(\mathbf{x})\) shows which pathway dominates~\cite{metzner2009transition}.

The challenge we set out to address is how one can use \(q^+(\mathbf{x})\) to describe the typical reaction mechanism when \(\mathbf{x}\) is too high dimensional to plot.
Because of the curse of dimensionality, we need to consider the components of \(\mathbf{x}\) one-by-one, inspecting how each one advances, and yet the two dimensional diffusion problem highlights the difficulty of decoupling those degrees of freedom.
A plot of \(q^+(x, y)\) contains information about how correlated \(x\) and \(y\) motion can conspire to advance the reaction.
If one were to simply neglect information about \(y\), one would lose those correlations.
Figure~\ref{fig:2ddiffusion} shows how our proposal, Eq.~\eqref{eq:main}, can one-by-one capture how each coordinate evolves, even resolving the multiple reaction pathways.
To highlight that capability, we solved for \(q^+(\mathbf{x})\) at inverse temperature \(\beta = 4\), a value at which both pathways meaningfully contribute to transitions.
As a function of reaction progress (\(q^+\)), we monitor how the distribution of each coordinate evolves, revealing distinctly bimodal distributions that form two channels in Fig.~\ref{fig:2ddiffusion}.
The plot in Fig.~\ref{fig:2ddiffusion}c, for example, reflects that progress can emerge either when \(y\) increases to a shallow intermediate basins around \((x,y) = (0, 1.5)\) or by holding \(y \approx 0\) and letting \(x\) do all of the work by climbing up and over a single saddle.
Because the plots show a \emph{distribution}, not just a mean at each value of \(q^+\), they contain rich information about how important the coordinate's motion is for enabling reaction progress.
The channels stretch along the vertical direction when reaction progress is relatively insensitive to the precise value of the coordinate, and they narrow when the coordinate is strongly driving the reaction.
For a two-dimensional \(\mathbf{x}\) we do not mean to suggest that plotting \(\rho^{\mathcal{AB}}(x, q^+)\) and \(\rho^{\mathcal{AB}}(y, q^+)\) is simpler to parse than a plot of \(q^+(x, y)\).
Rather, our point is that rich insights about reaction mechanism can be extracted by these collections of plots, which remain computable and interpretable even when the dimensionality grows.
We emphasize this point with the second example.

\subsection{Bistable switching in a seven-species gene toggle switch}
\label{sec:bistability}

\subsubsection{The GTS Model}

\begin{figure*}
    \centering
    \begin{tikzpicture}
        \node at (16.5,0.15) {\includegraphics[width=0.45\textwidth]{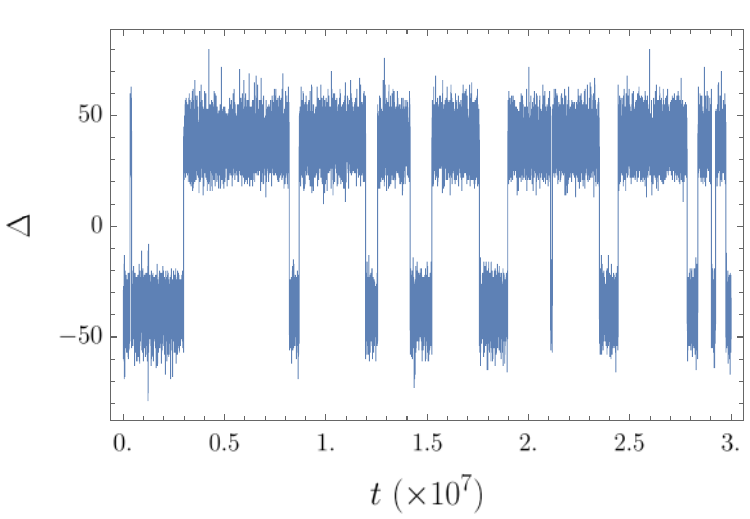}};
        \node at (5.7,2) {\ch{2 A <> [ \(k_1\) ][ \(k_2\) ] A_2}};
        \node at (5.6,1.1) {\ch{O + A_2 <> [ \(k_3\) ][ \(k_4\) ] OA_2}};
        \node at (6,0.2) {\ch{O -> [ \(k_5\) ] O + A}};
        \node at (6,-0.7) {\ch{OA_2 -> [ \(k_6\) ] OA_2 + A}};
        \node at (5.75,-1.6) {\ch{A -> [ \(k_7\) ] \(\varnothing\)}};
        \node at (9.2,2) {\ch{2 B <> [ \(k_1\) ][ \(k_2\) ] B_2}};
        \node at (9.1,1.1) {\ch{O + B_2 <> [ \(k_3\) ][ \(k_4\) ] OB_2}};
        \node at (9.5,0.2) {\ch{O -> [ \(k_5\) ] O + B}};
        \node at (9.5,-0.7) {\ch{OB_2 -> [ \(k_6\) ] OB_2 + B}};
        \node at (9.25,-1.6) {\ch{B -> [ \(k_7\) ] \(\varnothing\)}};
        \draw [draw=black] (11.2,2.6) rectangle (4,-2.1);
    \end{tikzpicture}
    \caption{\textbf{Gene toggle switch (GTS) model.}
      Left: Set of chemical reactions constituting the GTS model~\cite{allen2005sampling}, a chemical reaction network model with 7 species executing 14 stochastic reactions with 7 distinct reaction rates, \(k_1\) through \(k_7\).
      The system involves a single piece of DNA (species O), which can synthesize protein A or protein B.
      Those proteins can dimerize, and the dimers can bind to the DNA as a promoter that suppresses production of the other protein.
      The result is a bistable switch that toggles between an A-rich and B-rich state.
      Right: Which state the system occupies is well captured by the order parameter \(\Delta\) that counts how many more A proteins there are than B proteins (in monomer, dimer, or bound forms).
      A brute force stochastic simulation gives a Monte Carlo realization of a trajectory, illustrating the stochastic switching observed for the numerical values of the reaction rates reported in the main text.
      Fig.~\ref{fig:reaction} views this same process from the perspective of a statistical ensemble by employing transition path theory (TPT).
    }
    \label{fig:gts}
\end{figure*}
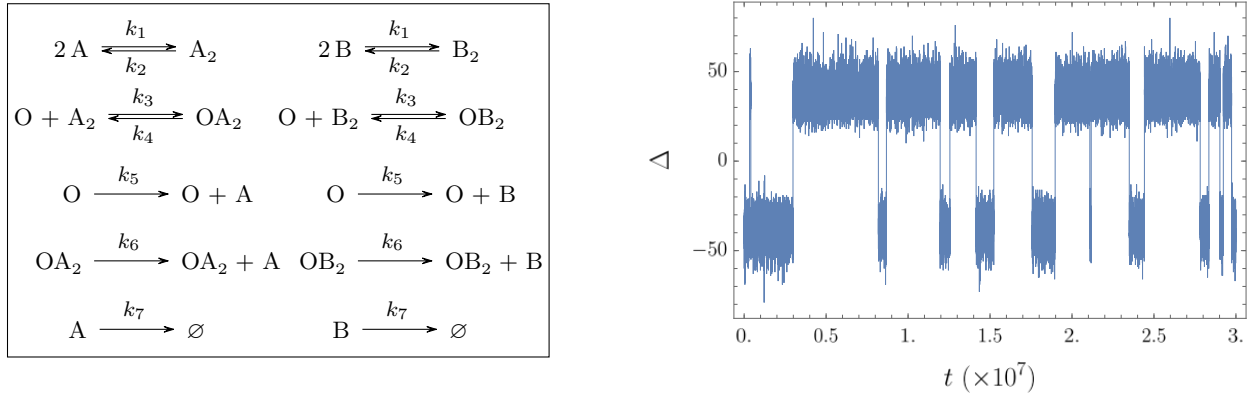

A paradigmatic example of bistable transitions in higher dimensions is provided by the chemical master equation (CME) for the stochastic dynamics of a gene toggle switch (GTS)~\cite{allen2005sampling}.
The GTS model we study was constructed to describe the fluctuating copy numbers of two proteins, A and B.
A single piece of DNA, denoted in the model by O, containing genes for A and B provides routes to increase the copy numbers through protein synthesis, but the copy numbers can also decrease via protein degradation.
The two genes mutually suppress each other, e.g., increasing the number of A decreases the production rate of B.
Consequently, the typical microstates involve either a high number of A or a high number of B, with rare stochastic fluctuations toggling between the metastable states.

The specific GTS model we study involves seven chemical species and fourteen reactions (see Fig.~\ref{fig:gts}).
The model allows for reversible dimerization of A and B to make A\(_2\) and B\(_2\).
Each dimer can also reversibly bind to the DNA to give OA\(_2\) and OB\(_2\).
The bound dimer acts as a promoter, so OA\(_2\) prompts the synthesis of more copies of protein A without similarly prompting the synthesis of B.
In the absence of a bound promoter, O is equally likely to synthesize A and B.
Finally, both proteins have an irreversible degradation process.
Figure~\ref{fig:gts} labels the rates for each of the fourteen elementary reactions by \(k_1\) through \(k_7\), assuming a symmetry between the kinetics for A and B. We follow Ref.~\cite{allen2005sampling}, setting \(k_1 = k_2 = k_3 = 5\), \(k_4 = k_5 = k_6 = 1\), and \(k_7 = 0.25\). The symmetry between A and B is spontaneously broken by the fluctuating dynamics, and the imbalance is monitored by the order parameter \(\Delta\equiv n_{\text A}+2 n_{\text A_2}+2 n_{\text{OA}_2}-n_{\text B}- 2 n_{\text B_2}-2 n_{\text{OB}_2}\), where \(n_\gamma\) denotes the number of species \(\gamma\). A microstate for the GTS model is then given by \((n_{\text A}, n_{\text{A}_2}, n_{\text{OA}_2}, n_{\text{O}},n_{\text{OB}_2}, n_{\text{B}_2},n_{\text{B}})\).
The representative stochastic trajectory of Fig.~\ref{fig:gts} shows that one can define a metastable \(\mathcal{A}\) region by \(\Delta \geq 25\) and a metastable \(\mathcal{B}\) region by \(\Delta \leq -25\).
Although the vast majority of the time is spent within either basin, we are primarily interested in the behavior of trajectories leaving \(\mathcal A\) and entering \(\mathcal B\).

Like the first example, the stochastic dynamics of the GTS is described by a Markovian jump process from one microstate to another.
The two-dimensional diffusion required a discretization onto a grid, but the states of the GTS naturally occupy a seven-dimensional lattice, one dimension per chemical species (A, A\(_2\), OA\(_2\), O, OB\(_2\), B\(_2\), and B).
None of the reactions can make or destroy the DNA, so there is a constant of motion: \(n_{\text{OA}_2} + n_{\text{O}} + n_{\text{OB}_2} = 1\).
That constraint restricts species OA\(_2\), O, and OB\(_2\) to each be present with either zero or one copy.
By contrast, the copy number of the A and B monomers and dimers can in principle increase without bound.
In practice, the degradation rate \(k_7\) ensures that there is some maximum copy number, \(M\), above which the dynamics is exceedingly unlikely to sample.
Appendix~\ref{sec:Rates} demonstrates that truncation at \(M = 30\) did not appreciably influence the reactive trajectories.
This choice of \(M\) is somewhat larger than what one might intuitively deduce from the positions of the distribution peaks as shown in Fig.~\ref{fig:rates}.
This is because the shape of the distribution, even on the level of typical events, is nontrivially influenced by rare configurations in the tails of the distribution.
Since we restrict \(n_{\text A}, n_{\text A_2}, n_{\text B},\) and \(n_{\text B_2}\) to each be between 0 and M, the state must be one of \(3 (M+1)^4\) microstates which comes to nearly 2.8 million for \(M=30\).
Though it is not entirely trivial to converge such a large vector, it is possible with Arnoldi and GMRES, particularly when there is a sparse representation of the Markov operator \(W\).

Provided the number of microstates is sufficiently modest that they can be practically enumerated, constructing the sparse matrix for \(W\) is straightforward.
For the \(i^{\rm th}\) microstate, one loops over the reactions in Fig.~\ref{fig:gts} to identify the microstate index \(j\) that would result if that reaction were to fire.
To the sparse matrix \(W\), one adds an element \(W_{ji} = \alpha\), where the so-called propensity \(\alpha\) is the rate of the reaction \(k_r\) times a combinatorial factor counting how many distinct copies of the species could have executed the reaction.
For example, if microstate \(i\) had 5 A monomers and 7 A\(_2\) dimers, then the reaction \(\ch{2 A -> [ \(k_1\) ] A_2}\) would have a rate \(20 k_1\) of mapping to the microstate with 3 A monomers and 8 A\(_2\) dimers.
Here, we have 20 distinct ways that two of the five A's could have participated in the reaction.
Any reaction that would have increased a copy number to exceed \(M\) would be disregarded.
Once all non-zero off-diagonal elements of \(W\) are identified, the diagonal elements are set to \(W_{ii} = -\sum_{j \neq i} W_{ij}\) to enforce conservation of probability.

The above procedure is conceptually simple and happens to be computationally tractable for this system size, but there is a more elegant way to build \(W\) that also extends to CMEs with astronomically large numbers of microstates.
The alternative approach leverages the Doi-Peliti (DP) formalism to represent \(W\) in terms of raising and lowering operators acting on each chemical species~\cite{doi1976stochastic,peliti1985path, nicholson2023quantifying}.
For the GTS problem, one can arrive at all of our results without the DP formalism, but we envision extensions to larger state spaces such that vectors like \({\bf q^+}\) cannot be explicitly computed but are rather approximated by a tensor network.
In those very large state spaces, looping over microstates is not possible and \(W\) must be built using the DP formalism.
Anticipating that necessity, we describe the DP representation of the GTS model in Appendix~\ref{sec:DP}.

\begin{figure}
    \centering
    \begin{tikzpicture}
        \node at (0.6,1.5) {\includegraphics[width=0.2\textwidth]{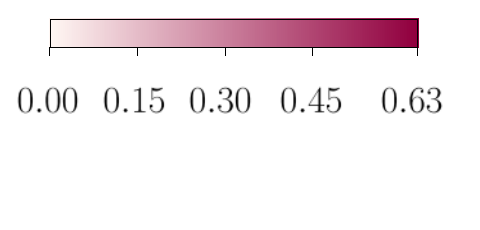}};
        \node at (0,0) {\includegraphics[width=0.4\textwidth]{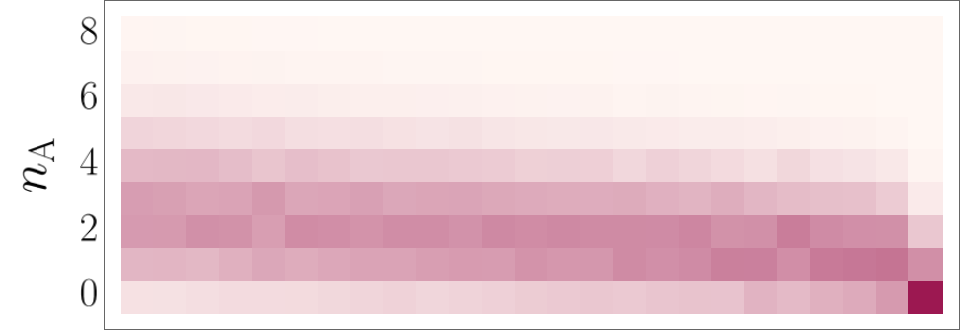}};
        \node at (-0.08,-3.3) {\includegraphics[width=0.409\textwidth]{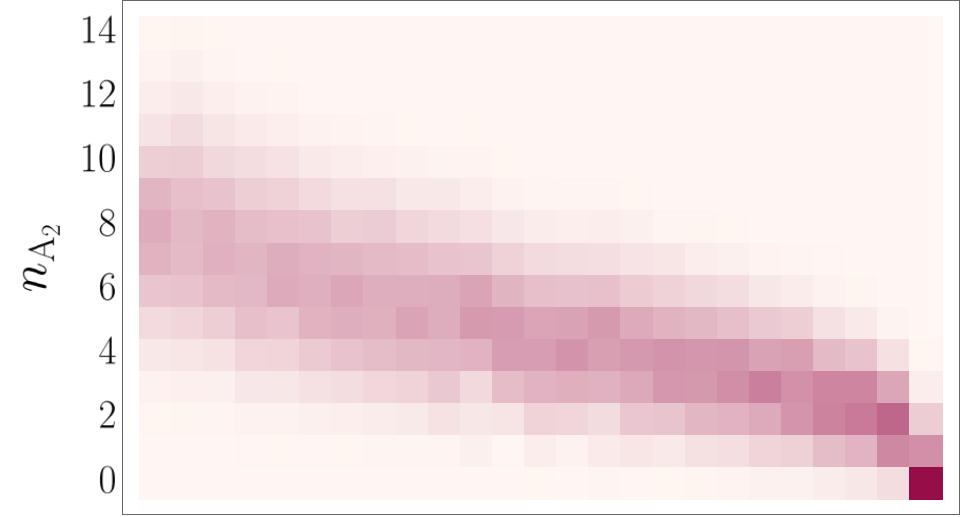}};
        \node at (-0.08,-7.3) {\includegraphics[width=0.409\textwidth]{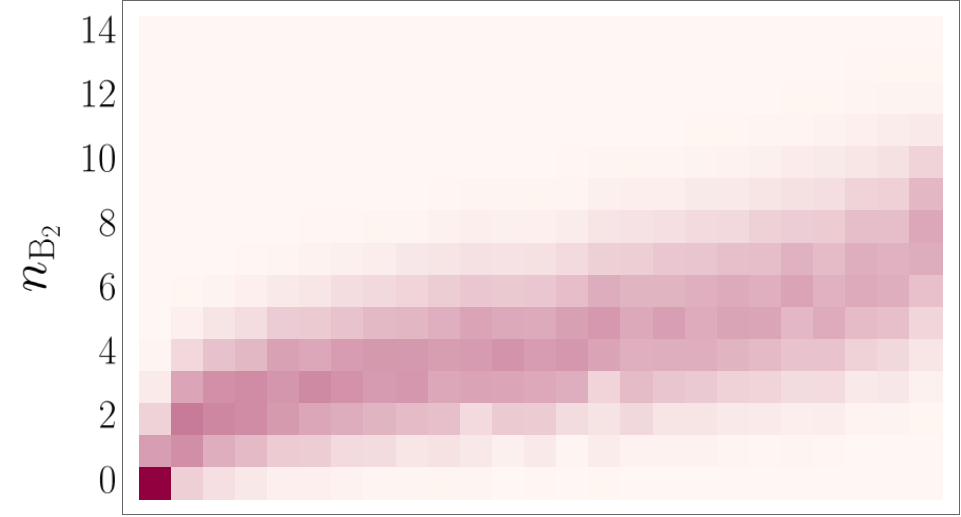}};
        \node at (0,-10.6) {\includegraphics[width=0.4\textwidth]{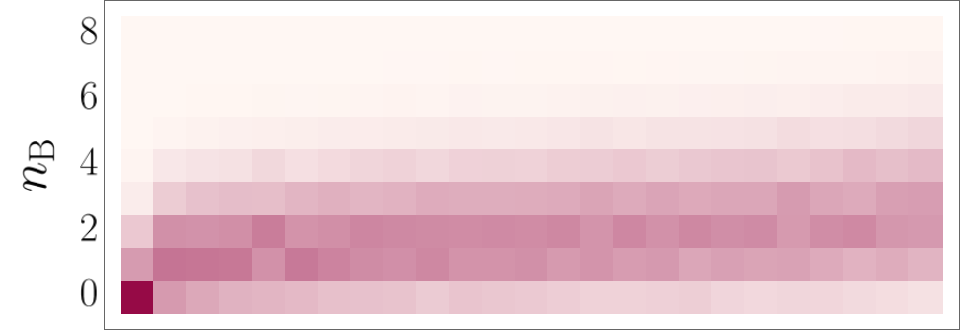}};
        \node at (-0.09,-14.3) {\includegraphics[width=0.414\textwidth]{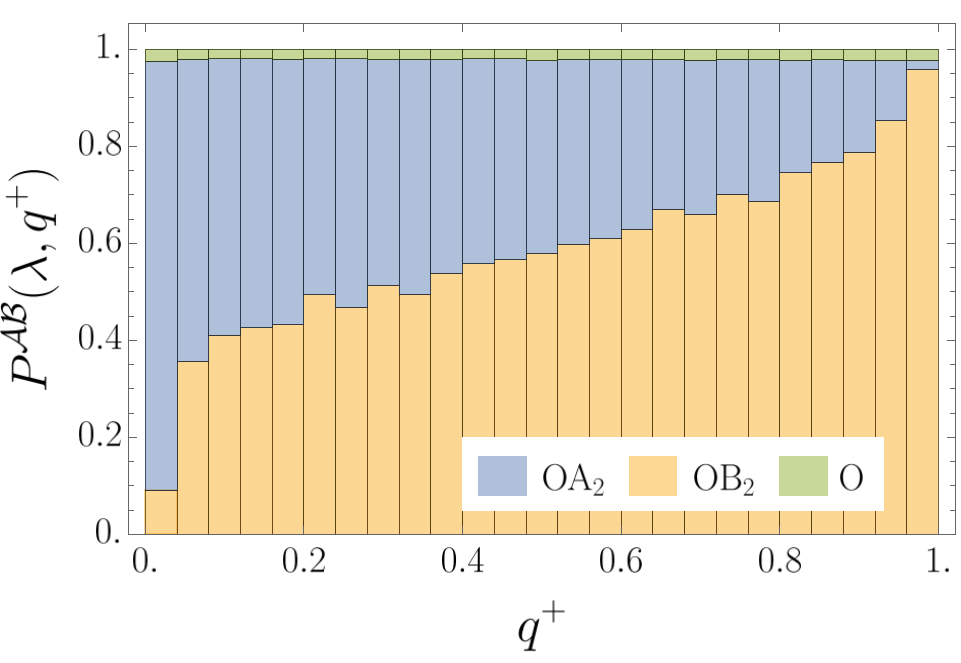}};
        \node at (-1.4,2.2) {\(P^\mathcal{AB}\)};
    \end{tikzpicture}
    \caption{\textbf{GTS reactive ensemble.}
      For the GTS model, the reaction coordinate \(q^+\) is explicitly computed with sparse linear algebra methods, allowing the distribution for each chemical species to be tracked as a function of reaction progress.
      Some species undergo significant changes in their average count, while others have more subtle changes in higher moments of their distribution.
      Counts of those chemical species are naturally discretized, but \(q^+\) is reported with bins of width 0.04 to aid in visualization.
      As marginals for the \emph{reactive} ensemble, these plots highlight typical reactive behavior moving from the edge of region \(\mathcal A\) to the edge of \(\mathcal B\).
      Configurations at those boundaries are not representative of typical steady-state behavior, which is dominated by \(\mathcal{A}\) and \(\mathcal{B}\).}
    \label{fig:reaction}
\end{figure}

\subsubsection{Analyzing the committor}
\label{sec:gtstrajectories}

Since the GTS model breaks detailed balance, we computed \(q^+(\mathbf{x}),\) and \(r(\mathbf{x})\) to obtain the reactive density \(P^{\mathcal{AB}}\) of Eq.~\eqref{eq:mAB}.
Fig.~\ref{fig:reaction} shows how the distribution for the number of each species evolves as a function of reaction progress \(q^+\).
Those distributions, \(P^{\mathcal{AB}}(n_\text{A}, q^+)\), \(P^{\mathcal{AB}}(n_{\text{A}_2}, q^+)\), \(P^{\mathcal{AB}}(n_\text{B}, q^+)\), and \(P^{\mathcal{AB}}(n_{\text{B}_2}, q^+)\), follow from the marginalizations of Eq.~\eqref{eq:main}, computed by discretizing \(q^+\) with bins of width 0.04.
The same procedure could also produce the reactive density for three different DNA states \(n_\text{O}, n_{\text{OA}_2},\) and \(n_{\text{OB}_2}\).
Since DNA must be in one and only one of those three states, it is more revealing to construct a new variable \(\lambda\) that records which of the three states the DNA is in.
That distribution over \(\lambda\) states then follows from a corresponding \(P^{\mathcal{AB}}(\lambda, q^+)\).

The five plots of Fig.~\ref{fig:reaction} collectively tell the story of how the elementary reactions of Fig.~\ref{fig:gts} collude together to allow the system to transit from A-rich to B-rich microstates.
Perhaps the clearest feature of the plots is the fact that the probability of finding the DNA in the O state is very small and completely insensitive to \(q^+\).
The calculations therefore show that the DNA will typically be bound to a dimer, and the reaction proceeds by switching from a bound \(\text{A}_2\) to a bound \(\text{B}_2\).
However, the reaction is not ``halfway done'' once the DNA flips from OA\(_2\) to OB\(_2\).
The bottom plot shows that \(q^+ \approx 0.35\) when OA\(_2\) and OB\(_2\) are equally likely in the reactive ensemble.
To push \(q^+\) beyond 0.5, it is also important that a sufficiently large population of B\(_2\) is built up, serving as a memory that prevents a rapid backsliding into the OA\(_2\) state.
Plots of \(P^{\mathcal{AB}}(n_{\text{B}}, q^+)\) and \(P^{\mathcal{AB}}(n_{\text{B}_2}, q^+)\) show that monomer and dimer play distinct roles.
DNA in the OB\(_2\) state produces only B, allowing for a buildup of monomer, but \(P^{\mathcal{AB}}(n_{\text{B}}, q^+)\) shows that the monomer does not appreciably build up over the course of the reaction.
While the distribution over \(n_\text{B}\) subtly shifts as a function of \(q^+\), it is always rare to see much more than 4 B molecules.
The relatively uniform fluctuations in \(n_\text{B}\) reflect that the number of B molecules is a poor proxy for the progress along the reaction coordinate. \(P^{\mathcal{AB}}(n_{\text{B}_2}, q^+)\) shows that it is instead the population of the dimer \(B_2\) that drives the progress to make the O\(\text{B}_2\) toward a stable B-rich state, a conclusion that follows from the drift in the peak of the \(n_{\text{B}_2}\) distribution as \(q^+\) grows.

\section{Discussion}

In this work, we have outlined and illustrated an approach to capturing the mechanism of transitions between two regions of very high-dimensional complex systems.
Our focus on rare events in noisy systems demands that we try to capture mechanism in a probabilistic way, seeking the evolution of the probability distribution for individual (physically interpretable) coordinates.
The first example emphasizes that these distributions need not be unimodal; there can be multiple dynamical pathways.
The rate operator \(W\) describes Markovian evolution of probability as a function of time, but that evolution superimposes transitions occurring at stochastically variable times.
TPT deconvolves that superposition, allowing us to resolve how the probability distribution over microstates evolves as a function of reaction progress, \(q^+\).
Our proposed marginalization of the reactive ensemble benefits from being straightforward and simple.
Simple, that is, provided the committor can be solved.

In this work, we solved for that committor for state spaces with millions of microstates, explicitly representing the vector \(\mathbf{q^+}\) and using sparse linear algebra methods to optimize it.
Those particular methods for finding the committor require that the state space be small enough that \(\mathbf{q^+}\) can be practically stored in memory, yet state spaces typically explode due to the many-body problem.
Even when that committor cannot be listed as a vector, dimensionality reduction strategies can robustly optimize the committor function.
For example, tensor train and tensor network approaches can extend sparse linear algebra methods to practically calculate properties of CMEs~\cite{hegland2010numerical,kazeev2014direct,liao2015tensor,dolgov2015simultaneous,gelss2016solving, vo2017adaptive,ion2021tensor}, including rare events for large (\(\sim10^{15}\) microstates) reaction-diffusion models~\cite{nicholson2023quantifying}.
Other approaches using basis expansions~\cite{thiede2019galerkin,strahan2021long,aristoff2024fast,guo2024dynamics}, neural networks~\cite{rotskoff2022active,strahan2023predicting,ma2005,khoo2018solving,hasyim2022supervised,li2019computing, jacques2023data,lin2024deep} or tensor networks~\cite{chen2023committor} can even fit high-dimensional committors for the case that \(\mathbf{x}\) is continuous.
As those strategies develop, it becomes especially important that one can use that committor function to extract information beyond a reaction rate.
We expect marginalizations of the reactive ensemble of the variety we describe to play an important role.

Those marginalizations may also be performed over transformed coordinates.
For our examples, we chose to work in a natural set of coordinates captured by \((x, y)\) and \((n_{\text A}, n_{\text{A}_2}, n_{\text{OA}_2}, n_{\text{O}},n_{\text{OB}_2}, n_{\text{B}_2},n_{\text{B}})\).
Our assumption is that these coordinates are simple to physically interpret as positions or counts of particular species.
We could alternatively consider correlations between the committor and some transformed physical coordinates.
For the first example, one could, for example, perform a rotation to study how distributions for the coordinates \((x+y, x-y)\) evolve with \(q\).
In principle, one would not be limited to rotations or linear transformations, but the more complex the transformation, the closer one gets to building \(q\) itself.
With a sufficiently complex transformation, one loses the physical interpretability and lands on the tautology ``the reaction proceeds because the reaction proceeds''.
To preserve mechanistic insights like ``the reaction proceeds because the distance between \(x\) and \(y\) shrinks'', it will be important to restrict ourselves to easily interpretable coordinates (the number of species A, the length of a bond, the number of solvent molecules within a radius of a protein, etc.).
Then Eq.~\eqref{eq:main} mixes the benefits of the two types of coordinates, revealing the correlations between simple-to-interpret physical coordinates and the simple-to-interpret concept of reaction progress.

\section{Acknowledgments}
We appreciate very useful discussions with Geyao Gu, Emanuele Penocchio, Aaron Dinner, Grant Rotskoff, Spencer Guo, and John Strahan.
The material presented in this manuscript is based upon work supported by the National Science Foundation under Grant No.\ 2239867.

\appendix
\section{Finite truncation}
\label{sec:Rates}

\begin{figure}[th]
  \centering
   \begin{tikzpicture}
       \node at (0,0) {\includegraphics[width=0.45\textwidth]{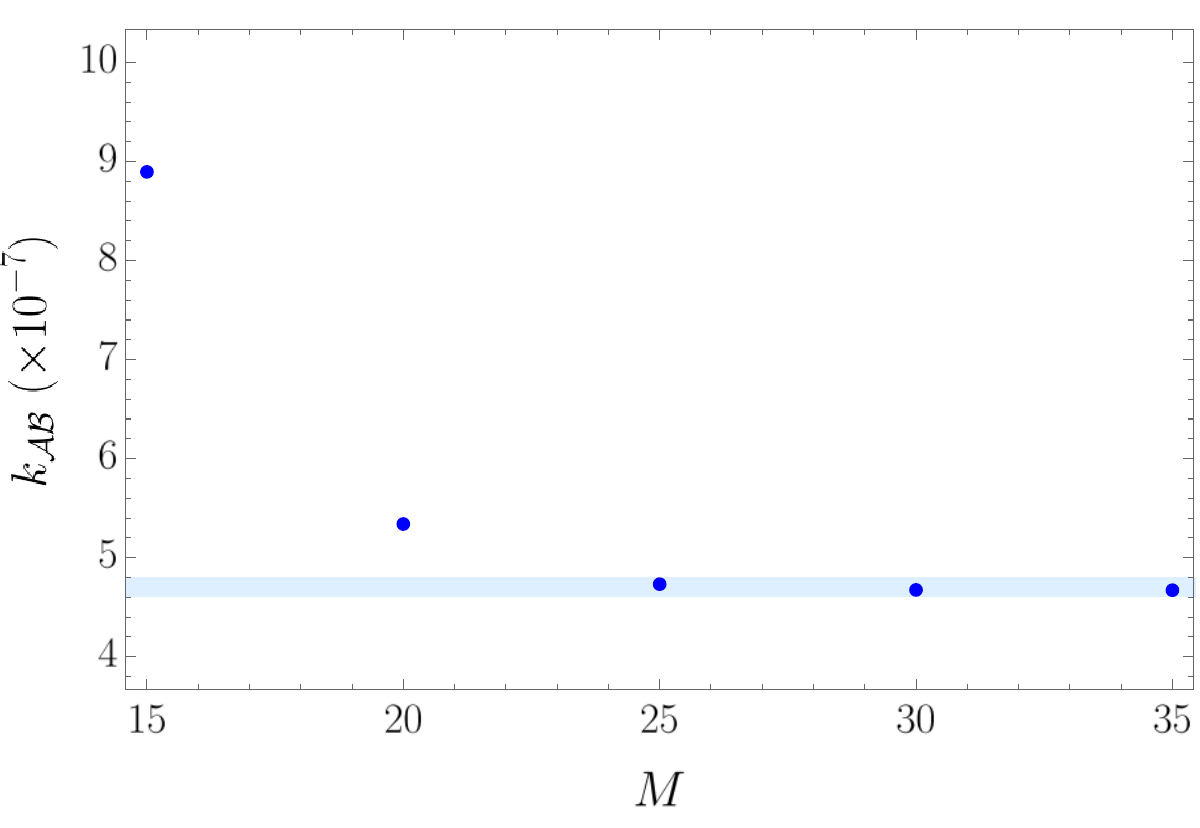}};
       \node at (1.2,0.7) {\includegraphics[width=0.3\textwidth]{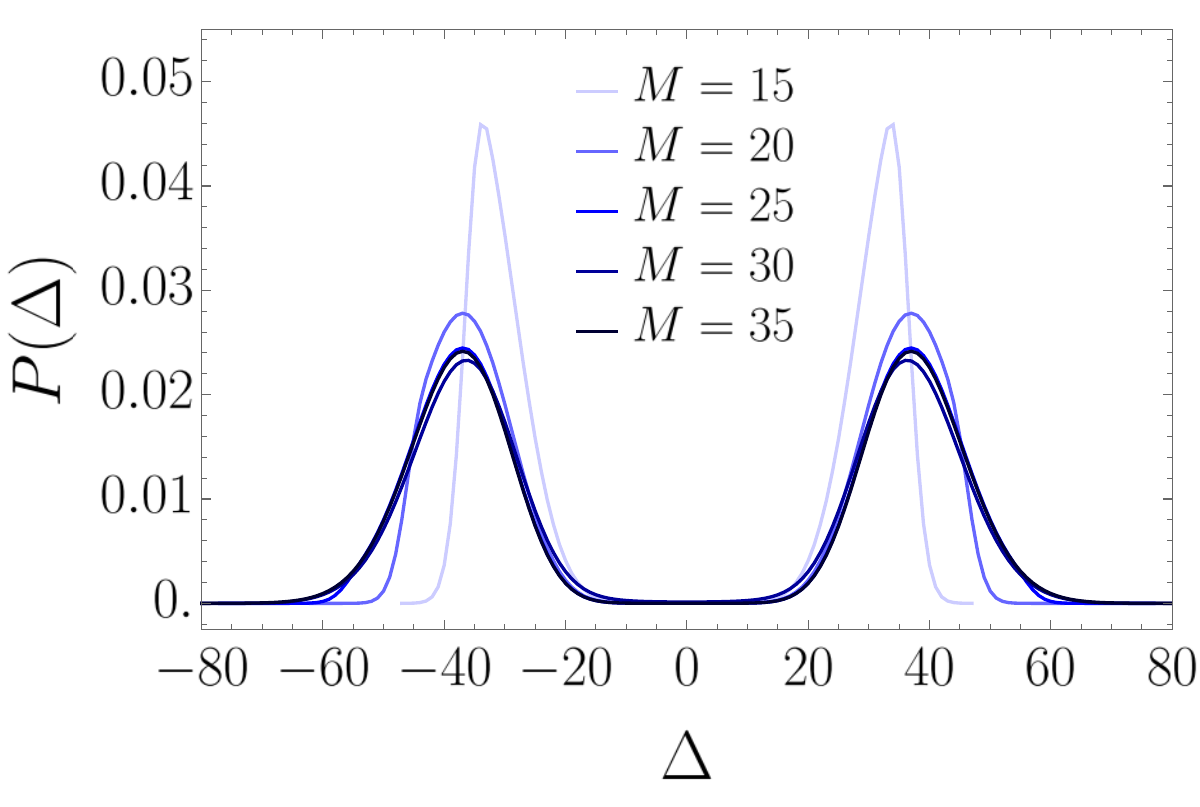}};
   \end{tikzpicture}
    \caption{\textbf{Finite truncation convergence.} Switching rates \(k_{\mathcal{AB}}\) between the two GTS metastable states were computed from Eq.~\eqref{eq:rate} for various choices of maximum molecule count \(M\).
      Provided \(M\) is sufficiently large, the committor-based rate calculations agree with rates obtained by FFS~\cite{allen2005sampling}, given by lines with thickness matching the reported standard errors.
      (Inset) The steady-state distribution for the order parameter \(\Delta\) also converges for the same sufficiently large \(M\).
    }
    \label{fig:rates}
\end{figure}

Our ability to generate the marginal reactive ensemble distributions required that we could directly compute \({\bf q^+}\), something we did in both example problems with sparse linear algebra methods.
Even with those sparse methods, it is important that one can cap the state space to prevent \({\bf q^+}\) from growing too large.
For the GTS model, our imposition of a maximum occupancy, \(M\) on non-DNA species, served this goal. In order to test that our cap set to \(M=30\) does not influence the reactive trajectories transitioning between \(\mathcal{A}\) and \(\mathcal{B}\), we compared rates computed via TPT with \(M = 30\) to rates computed via forward flux sampling (FFS)~\cite{allen2005sampling}, as well as the stochastic sampling algorithm (SSA) with no maximum occupancy~\cite{gillespie2013perspective}.
From TPT~\cite{metzner2009transition},
\begin{equation}
    k_{\mathcal{AB}}=\sum_{j\in\mathcal A,i\notin \mathcal A\cup\mathcal B}f_{ij}^{\mathcal{AB}},
    \label{eq:rate}
\end{equation}
where the flux of probability from microstate \(j\) to \(i\) within the reactive ensemble is
\begin{align}
    f_{ij}^{\mathcal{AB}}=\begin{cases}P_i^{\mathcal{AB}} W_{ij}\quad&i\neq j,\\
    0,\quad&\text{otherwise}\end{cases}.
    \label{eq:flux}
\end{align}
A truncation at \(M = 30\) was sufficient to yield \(k_{\mathcal{AB}} = 4.67 \times 10^{-7}\), a rate in excellent agreement with forward flux sampling (FFS) calculations performed on the same model~\cite{allen2005sampling} and a brute force rate calculation using 100 SSA trajectories, each of length \(10^8\) units of time.
Fig.~\ref{fig:rates} shows the convergence of the truncated TPT rates to \(k_{\mathcal{AB}} = (4.68\pm 0.05) \times 10^{-7}\), the SSA rate without a truncated maximal occupancy.

The inset to Fig.~\ref{fig:rates} emphasizes that \(M = 30\) was sufficient not only to converge the rate but also to converge distributions.
Specifically, we use \(\boldsymbol \pi\) to plot the steady-state distribution for \(\Delta\).
This \(P(\Delta)\), which reveals the bimodality for all \(M\), shows that (for the studied parameters) the distribution is only weakly influenced when \(M\) exceeds 25.

\section{Avoiding an ill-conditioned backward committor equation}
\label{sec:committors}
Here, we derive Eq.~\eqref{eq:backward2}, the linear equation that solves for \({\bf r}\) instead of the backward committor \({\bf q^-}\).
Observe from Eq.~\eqref{eq:mAB} that the reactive ensemble require a Hadamard product of \(\boldsymbol \pi\), \({\bf q^+}\), and \({\bf q^-}\).
Eq.~\eqref{eq:backward} is an ill-conditioned equation that would solve for \({\bf q^-}\), but we can convert it into a significantly better conditioned equation for \({\bf r}\), the Hadamard product of \({\boldsymbol \pi}\) and \({\bf q^-}\).
To see this conversion, we restrict ourselves to \(i, j \in (\mathcal{A} \cup \mathcal{B})^{\rm c}\) then substitute \(\tilde{U}_{ij} = \tilde{W}_{ji}\) and \(\tilde{v}_i = -\sum_{k \in \mathcal{A}} \tilde{W}_{ki}\) into Eq.~\eqref{eq:backward}:
\begin{equation}
  \sum_j\tilde W_{ji}q_j^-=-\sum_{k\in\mathcal A}\tilde W_{ki}.
\end{equation}
Rewriting this equation in terms of the time-forward matrix \(W\), we have
\begin{equation}
  \frac{1}{\pi_i} \sum_j W_{ij} \pi_j q_j^- = -\frac{1}{\pi_i}\sum_{k \in \mathcal{A}} W_{ik} \pi_k.
  \label{eq:wpis}
\end{equation}
For Eq.~\eqref{eq:wpis} hold for all \(i\), we therefore require
\begin{equation}
    \sum_j W_{ij} r_j=-\sum_{k\in\mathcal A}W_{ik}\pi_k.
\end{equation}
Finally, recalling that \(U\) and \(W\) are transposes of each other within the \((\mathcal{A} \cup \mathcal{B})^{\rm c}\) region, the expression simplifies to Eq.~\ref{eq:backward2}:
\begin{equation}
  U^{T} {\bf r} = {\bf s}.
\end{equation}

\section{Doi-Peliti construction of \(W\)}
\label{sec:DP}

A microstate of the GTS is given by \((n_{\text{A}}, n_{\text{A}_2}, n_{\text{OA}_2}, n_{\text{O}}, n_{\text{OB}_2}, n_{\text{B}_2}, n_{\text{B}})\).
Recognizing that the DNA exists in the \(\text{OA}_2, \text{O},\) or \(\text{OB}_2\) state, we equivalently define \({\bf n} = (n_{\text{A}}, n_{\text{A}_2}, n_{\lambda}, n_{\text{B}_2}, n_{\text{B}})\), where \(n_{\lambda} = 0, 1,\) and 2 correspond to the OA\(_2\), O, and OB\(_2\) states, respectively.
The vector of probabilities of each microstate, \({\bf p}_t\), evolves according to the master equation
\begin{equation}
  \frac{d \mathbf{p}_t}{d t} = W \mathbf{p}_t,
  \label{eq:mastereqn}
\end{equation}
where \(W\) is a rate operator constructed from the 14 reactions of Fig.~\ref{fig:gts}.
Writing that \(W\) in a matrix form can be an accounting headache that requires one to enumerate the microstates.
It can be convenient to instead write both \({\bf p}_t\) and \(W\) in a tensor-product form that isolates each reactions action on the occupation numbers of the chemical species.
Here, we sketch the framework for constructing \(W\) in terms of operators which raise and lower \(n_{\text{A}}, n_{\text{A}_2}, n_{\lambda}, n_{\text{B}_2},\) and \(n_{\text{B}}\).
Readers interested in more algebraic details are referred to the appendices of Ref.~\cite{nicholson2023quantifying}.

The aim is to write each reaction's contribution to the rate operator \(W\) in a tensor-product form:
\begin{equation}
O_{\text{A}} \otimes O_{\text{A}_2} \otimes O_{\lambda} \otimes O_{\text{B}_2} \otimes O_{\text{B}},
\end{equation}
where each \(O_\gamma\) is an operator that acts on a local state space spanned by the possible states of \(\left|n_\gamma\right>\).
Since A, A\(_2\), B, and B\(_2\) have an occupancy number between 0 and \(M\), their local state spaces are spanned by orthonormal basis vectors \(\left|0\right>, \left|1\right> \hdots \left|M\right>\), meaning the operators acting on their local state spaces are merely \((M+1) \times (M+1)\) matrices.
Operators on the \(\lambda\) space are even smaller \textemdash they are simply 3 \(\times\) 3 matrices.
Microstates of the many-body system are spanned by the tensor product states \(\left|{\bf n}\right> = \left|n_{\text{A}}\right> \otimes \left|n_{\text{A}_2}\right> \otimes \left|n_\lambda\right> \otimes \left|n_{\text{B}_2}\right> \otimes \left|n_{\text{B}}\right>\), which are also orthonormal.
We write a probability distribution over microstates as a superposition of the many-body states:
\begin{equation}
  \left|p_t\right> = \sum_{{\bf n}} p_t({\bf n}) \left|{\bf n}\right>.
  \label{eq:manybodyp}
\end{equation}
Eq.~\eqref{eq:manybodyp} is the tensor-product form of what we previously called \({\bf p}_t\).
By inspecting how each reaction impacts \(p_t({\bf n})\), we are now in a position to build the tensor-product form of \(W\).

To gain an intuition for how a chemical reaction gets converted into the set of local operators, it is useful to explicitly consider the first reaction of Fig.~\ref{fig:gts}, \(\ch{2 A -> [ \(k_1\) ] A_2}\).
The action of that reaction is to decrease \(n_{\text{A}}\) by two and to increase \(n_{\text{A}_2}\) by one, so it is useful to define a raising operator \(x_\gamma^\dagger\) and a corresponding lowering operator \(x_\gamma\) that act on species gamma.
Taking into account that species \(\gamma\) has a maximum occupancy of \(M_\gamma\), these operators are defined by
\begin{align}
    \nonumber x_\gamma|n_\gamma\rangle&=\begin{cases}n_\gamma|n_\gamma-1\rangle,\quad&0<n_\gamma\leq M_\gamma,\\
    0,\quad&\text{otherwise,}\end{cases}\\
    x^\dagger_\gamma|n_\gamma\rangle&=\begin{cases}|n_\gamma+1\rangle,\quad&0\leq n_\gamma<M_\gamma - 1.\\
    0,\quad&\text{otherwise.}\end{cases}
    \label{eq:ladder}
\end{align}
In matrix form,
\begin{equation}
\begin{split}
    x=\begin{pmatrix}
    0 & 1 & 0 & \cdots & 0 & 0\\
    0 & 0 & 2 & \cdots & 0 & 0\\
    0 & 0 & 0 & \cdots & 0 & 0\\
    \vdots & \vdots & \vdots & \ddots & \vdots & \vdots\\
    0 & 0 & 0 & \cdots & 0 & M\\
    0 & 0 & 0 & \cdots & 0 & 0\\
    \end{pmatrix}
\end{split}
\begin{split}
    \text{ and }
\end{split}
\begin{split}
    x^\dagger=\begin{pmatrix}
    0 & 0 & 0 & \cdots & 0 & 0\\
    1 & 0 & 0 & \cdots & 0 & 0\\
    0 & 1 & 0 & \cdots & 0 & 0\\
    \vdots & \vdots & \vdots & \ddots & \vdots & \vdots\\
    0 & 0 & 0 & \cdots & 0 & 0\\
    0 & 0 & 0 & \cdots & 1 & 0\\
    \end{pmatrix}.
\end{split}
\end{equation}
One might therefore guess that reaction 1 contributes to \(W\) a term of the form \(x_{\text{A}}^2 \otimes x^\dagger_{\text{A}_2} \otimes \mathbb{I}_{\lambda} \otimes \mathbb{I}_{\text{B}_2} \otimes \mathbb{I}_{\text{B}}\), a guess that involves lowering A two times, raising A\(_2\) once, and acting on the other species with the identity \(\mathbb{I}\) to leave them unchanged.
That guess correctly anticipates the off-diagonal components of \(W\), but to conserve probability, there is an additional negative element along the diagonal of \(W\).
That negative term is especially clear in the gain-loss CME for the first reaction:
\begin{align}
  \nonumber \frac{dp_t(\mathbf{n})}{dt} = k_1 \bigg[ & \left( n_{\text{A}} + 2 \right) \left( n_{\text{A}} + 1 \right) p_t(n_{\text{A}} + 2, n_{\text{A}_2} - 1) \nonumber \\
  & - n_{\text{A}} \left( n_{\text{A}} - 1 \right) p_t(n_{\text{A}}, n_{\text{A}_2}) \bigg]
  \label{eq:gainloss}
\end{align}

By summing both sides of Eq.~\eqref{eq:gainloss} over microstates \(\left(\text{i.e., }\sum_{\textbf{n}} \hdots \left|\mathbf{n}\right>\right)\) and by judiciously replacing terms like \(n_{\text{A}}\) by their number operator representation \(a^\dagger a\), the action of reaction 1 can be expressed as
\begin{align}
  \nonumber \frac{d \left|p_t\right>}{dt} = k_1 \bigg(&x_{\text{A}}^2 \otimes x^\dagger_{\text{A}_2} \otimes \mathbb{I}_{\lambda} \otimes \mathbb{I}_{\text{B}_2} \otimes \mathbb{I}_{\text{B}}\\
  &- x_{\text{A}}^{\dagger 2} x_{\text{A}_2}^2 \otimes y_{\text{A}_2} \otimes \mathbb{I}_{\lambda} \otimes \mathbb{I}_{\text{B}_2} \otimes \mathbb{I}_{\text{B}} \bigg) \left|p_t\right>,
  \label{eq:rxn1longform}
\end{align}
where
\begin{equation}
  y = \mathbb{I} - \left|M\right>\left<M\right| - \left|M-1\right>\left<M-1\right| = \begin{pmatrix}
    1 & \cdots & 0 & 0 & 0\\
    \vdots & \ddots & \vdots & \vdots & \vdots\\
    0 & \cdots & 1 & 0 & 0\\
    0 & \cdots & 0 & 0 & 0\\
    0 & \cdots & 0 & 0 & 0\\
  \end{pmatrix}
\end{equation}
adjusts the probability conserving diagonal element to accommodate for the fact that \(n_{\text{A}_2} = M+2 \to M\) transitions have been removed by the truncation~\cite{nicholson2023quantifying}.
To compress the notation, it is customary to suppress the identity operators and the tensor product symbols, writing reaction 1's contribution to \(W\) as simply
\begin{equation}
  W^{2 \text{A} \to \text{A}_2} = k_1\left(x_{\text{A}}^2 x_{\text{A}_2}^\dagger - x_{\text{A}}^{\dagger 2} x_{\text{A}}^2 y_{\text{A}_2}\right).
\end{equation}
Similar procedures can be carried out for the other 13 reactions.

For compactness, it is useful to additionally define
\begin{equation}
  z = \mathbb{I} - \left|M\right>\left<M\right| = \begin{pmatrix}
    1 & 0 & \cdots & 0 & 0\\
    0 & 1 & \cdots & 0 & 0\\
    \vdots & \vdots & \ddots & \vdots & \vdots\\
    0 & 0 & \cdots & 1 & 0\\
    0 & 0 & \cdots & 0 & 0\\
  \end{pmatrix}
\end{equation}
as well as a set of \(3 \times 3\) operators acting on the \(\lambda\) space:
\begin{align}
    \nonumber a^\dagger &=\begin{pmatrix}
    0 & 1 & 0\\
    0 & 0 & 0\\
    0 & 0 & 0\\
    \end{pmatrix},\ 
    b^\dagger=\begin{pmatrix}
    0 & 0 & 0\\
    0 & 0 & 0\\
    0 & 1 & 0\\
    \end{pmatrix},\\
    \nonumber a&=\begin{pmatrix}
    0 & 0 & 0\\
    1 & 0 & 0\\
    0 & 0 & 0\\
    \end{pmatrix},\ 
    b =\begin{pmatrix}
    0 & 0 & 0\\
    0 & 0 & 1\\
    0 & 0 & 0\\
    \end{pmatrix},\\
    \nonumber \alpha &=\begin{pmatrix}
    1 & 0 & 0\\
    0 & 0 & 0\\
    0 & 0 & 0\\
    \end{pmatrix},\ 
    \beta =\begin{pmatrix}
    0 & 0 & 0\\
    0 & 0 & 0\\
    0 & 0 & 1\\
    \end{pmatrix},\\
    \omega &=\begin{pmatrix}
    0 & 0 & 0\\
    0 & 1 & 0\\
    0 & 0 & 0\\
    \end{pmatrix}.
\end{align}
The operators \(a^\dagger\) and \(a\) respectively create and destroy OA\(_2\) from O while \(b^\dagger\) and \(b\) respectively create and destroy OB\(_2\).
The final three operators, \(\alpha, \beta,\) and \(\omega\) detect the occupancy of the OA\(_2\), OB\(_2\), and O states, respectively.
Having defined all the necessary local operators, we finally write down the contribution to \(W\) from each of the 14 reactions:
\begin{widetext}
  \begin{equation}
    \begin{split}
      \begin{aligned}
        \nonumber   W^{2 \text{A} \to \text{A}_2} &= k_1\left(x_{\text{A}}^2 x_{\text{A}_2}^\dagger - x_{\text{A}}^{\dagger 2} x_{\text{A}}^2 y_{\text{A}_2}\right)\\
        \nonumber W^{\text{A}_2 \to 2\text{A}} &= k_2\left(x_{\text{A}}^{\dagger 2} x_{\text{A}_2} - z_{\text{A}} x_{\text{A}_2}^\dagger x_{\text{A}_2}\right)\\
        \nonumber W^{\text O+\text A_2\rightarrow\text{OA}_2}&=k_3\left(x_{\text{A}_2} a^\dagger_{\lambda} -x_{\text{A}_2}^\dagger x_{\text{A}_2} \omega_\lambda \right)\\
        \nonumber W^{\text{OA}_2\rightarrow \text O+\text A_2}&=k_4\left(x_{\text{A}_2}^\dagger a_\lambda - y_{\text{A}_2}\alpha_\lambda\right)\\
        \nonumber W^{\text O\rightarrow\text O+\text A}&=k_5\left(x_{\text{A}}^\dagger \omega_\lambda - z_{\text{A}} \omega_\lambda\right)\\
        \nonumber W^{\text{OA}_2\rightarrow \text{OA}_2+\text A}&=k_6\left(x_{\text{A}}^\dagger \alpha_\lambda-y_{\text{A}}\alpha_\lambda\right)\\
        \nonumber W^{\text A\rightarrow\varnothing}&=k_7\left(x_\text{A}-x_{\text{A}}^\dagger x_{\text{A}}\right)
      \end{aligned}
    \end{split}
    \quad \quad
    \begin{split}
      \begin{aligned}
        \nonumber   W^{2 \text{B} \to \text{B}_2} &= k_1\left(x_{\text{B}_2}^\dagger x_{\text{B}}^2 - y_{\text{B}_2} x_{\text{B}}^{\dagger 2} x_{\text{B}}^2 \right)\\
        \nonumber W^{\text{B}_2 \to 2\text{B}} &= k_2\left( x_{\text{B}_2} x_{\text{B}}^{\dagger 2} -  x_{\text{B}_2}^\dagger x_{\text{B}_2} z_{\text{B}} \right)\\
        \nonumber W^{\text O+\text B_2\rightarrow\text{OB}_2}&=k_3\left(b^\dagger_{\lambda} x_{\text{B}_2} -\omega_\lambda x_{\text{B}_2}^\dagger x_{\text{B}_2}\right)\\
        \nonumber W^{\text{OB}_2\rightarrow \text O+\text B_2}&=k_4\left(b_\lambda x_{\text{B}_2}^\dagger - \beta_\lambda y_{\text{B}_2} \right)\\
        \nonumber W^{\text O\rightarrow\text O+\text B}&=k_5\left(
        \omega_\lambda x_{\text{B}}^\dagger - \omega_\lambda z_{\text{B}}\right)\\
        \nonumber W^{\text{OB}_2\rightarrow \text{OB}_2+\text B}&=k_6\left(\beta_\lambda x_{\text{B}}^\dagger - \beta_\lambda y_{\text{B}}\right)\\
        \nonumber W^{\text B\rightarrow\varnothing}&=k_7\left(x_\text{B}-x_{\text{B}}^\dagger x_{\text{B}}\right).
      \end{aligned}
    \end{split}
    \label{eq:dp}
  \end{equation}
\end{widetext}

Here, we have introduced this tensor-product form as a convenient way to construct \(W\) for sparse matrix operations, but we note that it is also the starting point to employ tensor network methods.
Those methods promise to make these committor calculations practical for even larger systems.

\bibliography{biblio.bib}

\end{document}